\documentclass[journal]{IEEEtran}
\usepackage{cite}
\usepackage{amsmath,amssymb,amsfonts}
\usepackage{graphicx}
\usepackage{algorithm}
\usepackage{algorithmic}
\usepackage{array}
\usepackage{textcomp}
\usepackage{booktabs}
\usepackage{multirow}
\usepackage{hyperref}
\hypersetup{hidelinks}

\usepackage[table,xcdraw]{xcolor} 
\usepackage{manyfoot}
\usepackage{pifont}
\usepackage{makecell,tabularx,array,adjustbox}

\newcolumntype{Y}{>{\centering\arraybackslash}p{0.073\columnwidth}}

\newcommand{\cmark}{\ding{51}}  
\usepackage{ragged2e,array}
\newcolumntype{P}[1]{>{\RaggedRight\arraybackslash}p{#1}}

\definecolor{critical}{RGB}{153, 51, 51}
\definecolor{high}{RGB}{204, 102, 102}
\definecolor{moderate}{RGB}{255, 204, 102}
\definecolor{low}{RGB}{153, 204, 153}
\definecolor{noinfo}{RGB}{239, 239, 239}

\begin{document}

\title{State of Abdominal CT Datasets: A Critical Review of Bias, Clinical Relevance, and Real-world Applicability}

\author{%
  Saeide~Danaei, %
  Zahra~Dehghanian, %
  Elahe~Meftah, %
  Nariman~Naderi, %
  Seyed~Amir~Ahmad~Safavi\mbox{-}Naini, %
  Faeze~Khorasanizade, %
  and Hamid~R.~Rabiee%
  \thanks{Manuscript received August 16, 2025; revised XXX. This work was supported by <funding, if any>. (Corresponding author: Hamid~R.~Rabiee.)}%
  \thanks{S.~Danaei, Z.~Dehghanian, S.~A.~A.~Safavi\mbox{-}Naini, and H.~R.~Rabiee are with the Data Science and Machine Learning Lab (DML), Department of Computer Engineering, Sharif University of Technology, Tehran, Iran (e-mail: rabiee@sharif.edu).}%
  \thanks{E.~Meftah and S.~A.~A.~Safavi\mbox{-}Naini are with Data-Driven and Digital Health (D3M), The Charles Bronfman Institute for Personalized Medicine, Icahn School of Medicine at Mount Sinai, New York, NY, USA.}%
  \thanks{N.~Naderi with Shahid Beheshti University of Medical Sciences, Tehran, Iran}
  \thanks{F.~Khorasanizade with Tehran University of Medical Sciences Cancer Research Institute, Tehran, Iran}
}

\markboth{IEEE Journal of Biomedical and Health Informatics,~Vol.~XX, No.~XX, August~2025}%
{Danaei \MakeLowercase{\textit{et al.}}: Abdominal CT Datasets Review}

\maketitle

\begin{abstract}
This systematic review critically evaluates publicly available abdominal CT datasets and their suitability for artificial intelligence (AI) applications in clinical settings. 
We examined 46 publicly available abdominal CT datasets (50{,}256 studies). Across all 46 datasets, we found substantial redundancy (59.1\% case reuse) and a Western/geographic skew (75.3\% from North America and Europe). A bias assessment was performed on the 19 datasets with \(\geq\)100 cases; within this subset, the most prevalent high-risk categories were domain shift (63\%) and selection bias (57\%), both of which may undermine model generalizability across diverse healthcare environments—particularly in resource-limited settings. To address these challenges, we propose targeted strategies for dataset improvement, including multi-institutional collaboration, adoption of standardized protocols, and deliberate inclusion of diverse patient populations and imaging technologies. These efforts are crucial in supporting the development of more equitable and clinically robust AI models for abdominal imaging.

\end{abstract}


\section{Introduction}
\IEEEPARstart{A}{bdominal} computed tomography (CT) imaging plays a pivotal role in modern diagnostic radiology, offering high-resolution views of critical abdominal organs, including the liver, pancreas, spleen, and kidneys\cite{abdominalimaging2020}. These images enable radiologists to diagnose diseases, monitor their progression, and support critical treatment decisions, including surgical planning. However, accurate interpretation of CT images requires specialized expertise, which is often limited. This scarcity can lead to diagnostic delays, particularly in rare or complex conditions, potentially affecting patient outcomes \cite{Thrall2018}. As a result, ongoing efforts to improve the efficiency and accuracy of medical imaging have spurred the development of advanced computational approaches \cite{Hosny2018}.

To address these challenges, artificial intelligence (AI) has emerged as a transformative tool in medical imaging. AI-driven models can assist in diagnosis, improve treatment planning, and generate enhanced visualizations—including three-dimensional reconstructions—that support surgical decision-making \cite{Loper2024}. AI systems also show promise for early disease detection, which is particularly valuable in emergency or resource-constrained environments. However, the success of AI in medical imaging is heavily dependent on the quality and diversity of the datasets used for training. Models trained on biased or limited datasets often fail to generalize across different clinical environments, reducing their reliability in practice \cite{Bell2024, Alabduljabbar2024}. Therefore, the availability of robust, well-annotated, and representative data is a prerequisite for developing trustworthy AI systems.

Abdominal organ segmentation plays a key role in extracting biomarkers and quantifying tumor burden, making dataset quality even more critical \cite{tang2019clinically}. Unfortunately, many available datasets suffer from inherent biases that limit their clinical applicability. Spectrum bias, for instance, arises when datasets disproportionately include certain pathologies while under-representing others, resulting in skewed model performance \cite{Kocak2024}. Similarly, selection bias can occur when datasets lack diversity in patient demographics or disease stages. These biases undermine the generalizability of AI models and complicate their deployment in real-world scenarios. As discussed in the methodology section, addressing these issues is essential to ensure that AI tools trained on these datasets can function reliably across varied clinical settings. In this regard, image quality, annotation fidelity, and metadata completeness must collectively reflect the diversity encountered in routine clinical practice.

For AI to achieve its full potential in abdominal imaging, datasets must be diverse, well-annotated, and minimally biased—goals that demand international collaboration. As the number of publicly available abdominal CT datasets continues to grow, a critical question arises: Are these datasets truly fit for clinical translation and AI-driven decision-making? \cite{ueda2024fairness} In this review, we systematically evaluate existing abdominal CT datasets in terms of imaging characteristics, annotation standards, demographic representation, and clinical relevance. By identifying key limitations and sources of bias, we aim to offer a roadmap for improving dataset quality and ensuring that AI in medical imaging equitably serves all patient populations.

\section{Methods}\label{sec4}

To ensure a thorough and objective review of publicly available abdominal CT datasets, we adopted a structured evaluation framework tailored to machine learning applications in medical imaging. Our methodology encompasses dataset identification, selection criteria, and a multi-dimensional assessment of dataset quality and bias. By systematically analyzing annotation quality, demographic representation, and imaging characteristics, we aim to highlight gaps and opportunities for future dataset improvements.

\subsection{Dataset Identification}\label{subsec2}
To compile a comprehensive list of publicly available abdominal CT datasets, we conducted a structured search using Google Scholar, PubMed\cite{pubmed_nlm}, Scopus\cite{scopus_elsevier}, and institutional repositories such as the NIH and The Cancer Imaging Archive (TCIA). (Tables \ref{tab:datasets_1} and \ref{tab:datasets_2} shows datasets details). These sources were chosen for their extensive coverage of medical imaging datasets. The search was conducted over one month and iteratively refined with co-author feedback, ensuring comprehensive dataset identification. The primary keywords utilized during the search included:

\begin{itemize}
\item CT Scan Dataset(s)
\item Annotated CT Dataset(s)
\item CT Image Segmentation Dataset(s)
\item Tumor Detection in CT Scans
\item Abdominal Organ Segmentation in CT
\end{itemize}

All extracted fields (dataset identifiers, centers, organs, labels, provenance, and bias calls) are provided in a public spreadsheet.
(\S\ Data and Materials Availability). We used that sheet as the single source of truth for all tables/figures.

\subsection{Inclusion criteria}\label{subsec2}
We established strict inclusion criteria to ensure dataset relevance and quality:

\begin{itemize}
\item Annotation Requirement: Datasets must include labeled regions (e.g., organ contours, tumor boundaries) in formats such as polygonal segmentation, bounding boxes, or voxel-based annotations.

\item Clinical Relevance: Included datasets must support organ segmentation, anomaly detection, or disease diagnosis tasks.

\item Focus on Abdominal Organs: Annotations must pertain to key abdominal structures (e.g., liver, kidneys, spleen, pancreas).

\item Scientific Validation: Each dataset must be referenced in at least one peer-reviewed study demonstrating its application in medical imaging or machine learning.
\end{itemize}

\subsection{Evaluation Metrics}\label{subsec3}
Each dataset was systematically evaluated based on the following key parameters:

\begin{itemize}
    \item \textbf{Dataset Composition}: We recorded the number of publicly available vs. private CT studies, assessing dataset growth over time.
    
    \item \textbf{Case Status}: The clinical context of each dataset (e.g., disease presence, healthy controls) was analyzed to determine its applicability to real-world diagnosis.
    
    \item \textbf{Data Provenance}: We identified data sources, contributing institutions, and geographic representation to assess dataset diversity.
    
    \item \textbf{Imaging Quality}: We examined scan resolution, number of slices per study, and imaging protocols to evaluate dataset granularity.
    
    \item \textbf{Annotation Details}: We analyzed annotated organs, segmentation techniques, and labeling consistency to determine dataset reliability.
    
    \item \textbf{Demographic Diversity}: We reviewed available metadata on patient age, sex, and geographic distribution to gauge representational bias.
\end{itemize}
To complement these quantitative assessments, we conducted a bias evaluation to better understand how dataset characteristics influence fairness and generalizability across diverse clinical settings.


\subsection{Evaluation of Bias and Relevance}\label{subsec4}
To rigorously assess dataset fairness and representational validity, we performed a comprehensive bias evaluation on datasets with over 100 cases. Smaller datasets were excluded due to their high variance and limited statistical power, which diminishes the reliability of bias estimation \cite{Collins2024}. Our analysis covered eight distinct bias categories, each assessed independently despite some conceptual overlap. This method provided a detailed understanding of dataset limitations and their implications for real-world clinical applicability \cite{Kocak2024}.

We enlisted a qualified medical doctor with experience in both coding and the academic aspects of computer vision and machine learning, particularly in medical imaging, and who has previously worked with imaging datasets to conduct a systematic review of each dataset using our predefined bias assessment framework, which evaluated the following aspects:

\begin{itemize}
\item \textbf{Spectrum Bias}: Over-representation of specific conditions, potentially skewing model performance.
\item \textbf{Selection Bias}: Limited case diversity, impacting the model’s ability to generalize.
\item \textbf{Racial (or Ethnic) Bias}: Under-representation of specific racial or ethnic groups, leading to reduced model performance for those populations.
\item \textbf{Geographical (Developing World) Bias}: Dataset imbalances based on region, affecting disease prevalence representation and imaging protocol consistency.
\item \textbf{Technical Bias  (Protocol Bias)}: Variations in imaging protocols affecting data consistency.
\item \textbf{Labeling Bias (Annotation or Observer Bias)}: Annotation discrepancies due to differing expert guidelines.
\item \textbf{Temporal Bias}: Changes in imaging technology or clinical practices over time.
\item \textbf{Domain Shift Bias}: Performance inconsistencies when models are applied to external datasets.
\end{itemize}

In developing our evaluation protocol, we prioritized primary documentation sources to ensure maximum accuracy. For standalone datasets, we relied on published data descriptors as the principal reference, while for composite datasets, we systematically examined constituent dataset papers. In cases where formal descriptors were unavailable, we analyzed submission documentation from dataset repositories as the authoritative source.

In our evaluation, each dataset received a three-tier classification (low, medium, or high risk) for each bias type. When datasets lacked sufficient information for a given bias category, they were marked as not provided.

To derive an overall bias classification, we employed a structured scoring system:
\begin{itemize}
    \item \textbf{Critical Bias}: Datasets with five or more high-risk bias indicators.
    \item \textbf{High Bias}: Datasets with three to four high-risk bias indicators.
    \item \textbf{Low Bias}: Datasets with five or more low-risk bias indicators.
    \item \textbf{Medium Bias}: All remaining datasets.
\end{itemize}
To account for the impact of moderate and missing bias assessments, we implemented an equivalence formula\cite{Kocak2024}:
\begin{itemize}
    \item Every three medium-risk classifications were weighted as equivalent to one high-risk field.
    \item Every two not-provided fields were similarly weighted as one high-risk field.
\end{itemize}

To ensure methodological consistency and reliability, the bias reviewer underwent preparatory training on bias identification, classification, and evaluation standards. This training was critical for minimizing inter-rater variability and ensuring a standardized approach across all assessments.

Bias assessments were systematically documented in a structured matrix and incorporated into our analytical framework to enable standardized comparisons. This methodology provided essential context for dataset quality assessment and facilitated standardized comparisons of strengths and limitations across datasets. By embedding bias evaluation into our primary dataset characterization, we ensured that bias-related limitations informed subsequent analysis phases and interpretation of findings.

\subsection{Evaluation of Adaptability in Developing Countries}\label{subsec5}

Given that most abdominal CT datasets originate from high-resource settings, we assessed their applicability in developing countries using three key factors:
\begin{itemize}
    \item \textbf {Geographic Diversity}: We examined whether datasets included scans from low- and middle-income countries.
    \item \textbf {Demographic Representativeness}: We analyzed patient populations to ensure diverse age, sex, and ethnic distributions.
    \item \textbf {Technological Compatibility}: We prioritized datasets including older-generation CT scanners, which are commonly used in resource-limited hospitals.
\end{itemize}

Ensuring the cross-contextual validity of abdominal CT datasets is critical for their applicability in resource-constrained healthcare environments. Since most publicly available datasets originate from high-resource settings, their generalizability to low- and middle-income countries (LMICs) remains uncertain, given differences in clinical practices, imaging technologies, and patient demographics.

To systematically assess dataset adaptability, we evaluated three key factors:
\begin{itemize}
    \item Geographical Diversity of Data Sources: The extent to which datasets include CT scans from non-Western regions or multi-center contributions spanning diverse healthcare settings.
    \item Demographic Representativeness: The balance of age, sex, ethnicity, and socioeconomic factors within patient populations, ensuring fair representation across global populations.
    \item Technological Heterogeneity: The inclusion of scans acquired from older-generation CT scanners, which remain widely used in developing countries, makes such datasets more relevant for real-world applications.
\end{itemize}

Given the significant technological disparities between high-resource and resource-limited settings, we prioritized datasets containing scans from older CT models, as they better reflect the imaging infrastructure in many hospitals and diagnostic centers worldwide.

The overarching goal of this evaluation framework was to identify datasets with strong cross-contextual applicability—those capable of supporting AI-driven solutions that remain robust and diagnostically useful across diverse clinical environments. Without such adaptability, AI models trained on Western-centric datasets may struggle to generalize in LMICs, exacerbating health disparities rather than alleviating them.

By highlighting these limitations, our analysis underscores the urgent need for more inclusive dataset curation, with deliberate efforts to incorporate data from underrepresented regions with different imaging technologies. Without such measures, the full potential of AI-driven medical imaging cannot be realized on a truly global scale.

\section{Results}\label{sec2}

\subsection{Overview of Datasets}\label{subsec1}
Based on the structured evaluation framework outlined in the Methodology, this section presents key findings regarding the composition, annotation practices, dataset bias, and demographic diversity of publicly available abdominal CT datasets. We analyzed 46 datasets encompassing a total of 50,256 CT studies to assess their suitability for AI-driven medical applications. Tables \ref{tab:datasets_1} and \ref{tab:datasets_2} indicate summarized details such as the number of volumes, the proportion of cases reused, pathology status, contributing centers, source countries, annotated organs, the availability of anomaly labels, and annotation methods.

\begin{table*}[ht]
    \centering
    \caption{Dataset Volume and Subject Information} \label{tab:datasets_1}
    \footnotesize
    \setlength{\tabcolsep}{4pt}
    \renewcommand{\arraystretch}{1.1}
    \begin{tabular}{|m{5cm}|m{2cm}|m{4cm}|m{6cm}|}
        \hline
        \textbf{Dataset Name} & \textbf{cases \footnotemark[1]} & \textbf{Reused Cases} & \textbf{Subjects Status} \\
        \hline
        SLIVER (\href{https://sliver07.grand-challenge.org}{2007})\cite{sliver} & 20+10 & 0 & Most cases had tumors, metastasis, and cysts of different sizes    \\
        3D-IRCADb (\href{https://www.ircad.fr/research/data-sets/}{2010}) & 22+0 & 0 & Liver tumors, FNH cases. \\
        VISCERAL (\href{https://visceral.eu/benchmarks/}{2015}) \cite{viserectal} & 40+27 & 0 & "bone marrow" neoplasms \\
        BTCV (\href{https://www.synapse.org/Synapse:syn3193805/wiki/217789}{2015}) \cite{btcv2015} & 30+20 & 0 & Cancer, post-op hernia. \\
        Colorectal-Liver-Metastases(\href{https://www.cancerimagingarchive.net/collection/colorectal-liver-metastases/}{2017}) \cite{simpson2024preoperative} & 394+0 & 0 & CRC with liver metastases. \\
        DenseVNet (\href{https://zenodo.org/records/1169361}{2018}) \cite{densevnet} & 90+0 & 100\% (\cite{btcv2015}\cite{roth2016pancreasct}) & Healthy, liver metastases. \\
        LiTS (\href{https://academictorrents.com/details/27772adef6f563a1ecc0ae19a528b956e6c803ce}{2018}) \cite{Lits2017} & 131+70 & 9.95\% (\cite{3D-IRCADb-01}) & Liver cancer, pre/post-therapy \\
        MSD-CT - Spleen task (\href{http://medicaldecathlon.com/}{2018}) \cite{msd2022} & 41+20 & 0 & Liver metastases post-chemo \\
        Pancreatic Cancer Survival Prediction (\href{https://wiki.cancerimagingarchive.net/pages/viewpage.action?pageId=37224869}{2018}) & 159+53 & 0 & Candidates for pancreatic cancer resection \\
        MSD-CT - Colon task (\href{http://medicaldecathlon.com/}{2018}) \cite{msd2022} & 126+64 & 0 & Candidates for colorectal cancer resection \\
        SegThor (\href{https://competitions.codalab.org/competitions/21145}{2019}) \cite{segthor} & 40+20 & 0 & NSCLC, curative radiotherapy \\
        CHAOS (\href{https://chaos.grand-challenge.org/Data/}{2019}) \cite{CHAOS2021} & 20+20 & 0 & Healthy donors, atypical livers \\
        CT-ORG (\href{https://www.cancerimagingarchive.net/collection/ct-org/}{2020}) \cite{ctorg2020} & 119+21 & 93.57\% (\cite{Lits2017}) & liver lesions (benign/malignant) with cancers of other organs \\
        MSD-CT - Pancreas task (\href{http://medicaldecathlon.com/}{2020}) \cite{msd2022} & 281+139 & 0 & Candidates for pancreatic mass resection \\
        MSD-CT - Liver task (\href{http://medicaldecathlon.com/}{2020}) \cite{msd2022} & 131+70 & 100\% (\cite{Lits2017}) & Liver cancer, pre/post-therapy \\
        MSD-CT - HepaticVessel task (\href{http://medicaldecathlon.com/}{2020}) \cite{msd2022} & 303+140 & 0 & Primary, metastatic liver tumors \\
        Pancreas-CT (\href{https://www.cancerimagingarchive.net/collection/pancreas-ct/}{2020})\cite{roth2016pancreasct} & 80+0 & 0 & Healthy donors, non-pancreatic cases \\
        AbdomenCT-1K (\href{https://doi.org/10.5281/zenodo.5903099}{2021})\cite{abdomenct2021} & 1112+0 & 95.5\%(multiple datasets\footnotemark[4]) & Various abdominal cancers \\
        WORC - GIST dataset (\href{https://zenodo.org/records/5221034}{2021}) \cite{Starmans2021WORC} & 246+0 & 0 & GIST, intra-abdominal tumors resembling GIST \\
        WORC - CRLM dataset (\href{https://zenodo.org/records/5221034}{2021})\cite{worcdatabase} & 77+0 & 0 & CRC liver metastases \\
        Pediatric (\href{https://www.cancerimagingarchive.net/collection/pediatric-ct-seg/}{2022})\cite{pediatric} & 359+0 & 0 & Pediatric CT cases \\
        WORD (\href{https://drive.google.com/drive/folders/1i2xbXxdEYnjNZVUtGZxYdwaeKmNmywnY}{2022}) \cite{word2022} & 170+0 & 11.76\% (\cite{Lits2017}) & Cancer, pre-radiotherapy \\
        AMOS (\href{https://zenodo.org/records/7262581}{2022})\cite{amos} & 500+0 & 0 &  Abdominal tumors, other nonmalignant abdominal pathologies \\
        KiPA22 (\href{https://zenodo.org/records/6361938}{2022}) & 100+30 & 0 &  Renal tumors affecting only one kidney \\
        StageII-Colorectal-CT (\href{https://www.cancerimagingarchive.net/collection/stageii-colorectal-ct/}{2022}) \cite{stageii_crc} & 230+0 & 0 & Stage II CRC, pre-op CTs \\
        HCC-TACE-Seg (\href{https://www.cancerimagingarchive.net/collection/hcc-tace-seg/}{2022}) \cite{moawad2021hcc} & 211+0 & 0 & HCC, TACE treatment cases \\
        
        AutoPET (FDG-PET/CT)(\href{https://www.cancerimagingarchive.net/collection/fdg-pet-ct-lesions/}{2022}) \cite{automatedpet} & 1014+150 & 0 &  Oncological cases (mostly NSCLC, lymphoma, melanoma) \\
        
        DAP Atlas (\href{https://www.synapse.org/Synapse:syn52287632.1/datasets/}{2023}) \cite{dapatlas2023} & 533+0 & 100\% (\cite{automatedpet}) & cancer, tumors, and enlarged anatomical structures \\
        
        Abdominal Trauma Det (\href{https://www.rsna.org/rsnai/ai-image-challenge/abdominal-trauma-detection-ai-challenge}{2024}) \cite{rudie2024rsna} & 3551+723 & 0 & Traumatic injuries \\
        
        TotalSegmentator (\href{https://zenodo.org/records/10047292}{2023}) \cite{totalsegmentator} & 1204+0 & 0 & Mixed normal/pathology \\

        AbdomenAtlas 1.1 (\href{https://www.zongweiz.com/dataset}{2024}) \cite{abdomenatlas2024}& 9262+11223 & 60.02\%(multiple datasets\footnotemark[5]) & Normal and cancerous organs (colorectal, pancreatic)\\
    
        FLARE23 (\href{https://codalab.lisn.upsaclay.fr/competitions/12239}{2023}) \cite{flare23} & 4250+400 & 100\% (multiple datasets\footnotemark[6] & Normal and cancerous cases\\
        
        KiTS (\href{https://kits-challenge.org/kits23/}{2019}) \cite{kits2019} & 489+110 & 0 & kidney tumor or cysts suspicious of malignancy\\
        
        CPTAC-PDA-Tumor-Annotations (\href{https://www.cancerimagingarchive.net/analysis-result/cptac-pda-tumor-annotations/}{2023})\cite{rozenfeld2023cptac} & 97+0 & 0 & Pancreatic ductal adenocarcinoma \\
        
        CPTAC-CCRCC-Tumor-Annotations (\href{https://wiki.cancerimagingarchive.net/pages/viewpage.action?pageId=157288300}{2023})\cite{clark2013tcia} & 55+0 & 0 & lear Cell Renal Cell Carcinoma, pre/post-treatment\\
        
        CARE (\href{https://paperswithcode.com/dataset/care}{2023}) \cite{zhang2023care}& 399+0 & 0 & rectal cancer and its surrounding normal tissue\\
        
        Low-dose (\href{https://www.cancerimagingarchive.net/collection/ldct-and-projection-data/}{2023})\cite{moen2021lowdose} & 75+0 & 0 & liver metastasis \\
        
        SEG.A. (\href{https://multicenteraorta.grand-challenge.org/}{2023}) & 56+0 & 100\% (\cite{kits2019}) & aortic pathologies\\
        
        CT Lymph Nodes (\href{https://www.cancerimagingarchive.net/collection/ct-lymph-nodes/}{2023})\cite{roth2014lymphnode} & 86+0 & 0 & Non-cancerous lymphadenopathy\\
        
        Adrenal-ACC-Ki67-Seg (\href{https://www.cancerimagingarchive.net/collection/adrenal-acc-ki67-seg/}{2023})\cite{ahmed2020radiomic} & 65+0 & 0 & Adrenocortical carcinoma with assessed Ki-67 index\\
        
        AIMI Annotations Initiative (\href{https://zenodo.org/records/13244892}{2024}) \cite{vanoss2024aimi} & 1231+0 & 100\% (\cite{moawad2021hcc}\cite{simpson2024preoperative}\cite{clark2013tcia}) & kidney and liver tumor\\
        
        CURVAS (\href{https://zenodo.org/records/10979642}{2024}) \cite{riera2024curvas} & 20+70 & 0 & cysts and other pathologies (benign and malignant)\\
        
        \hline
    \end{tabular}
\end{table*}

\begin{table*}[ht]
    \centering
    \caption{Dataset Volume and Subject Information} \label{tab:datasets_2}
    \footnotesize
    \setlength{\tabcolsep}{4pt}
    \renewcommand{\arraystretch}{1.1}
    \begin{tabular}{|m{5cm}|m{3cm}|m{4cm}|m{2cm}|m{3cm}|}
        \hline
        \textbf{Dataset Name} & \textbf{Centers \footnotemark[2]} & \textbf{Annotated Organs \footnotemark[3]} & \textbf{Anomaly label} & \textbf{Annotation method} \\
        \hline
    SLIVER (\href{https://sliver07.grand-challenge.org}{2007})\cite{sliver} & -- & L & -- & expert 
    \\
    
    3D-IRCADb (\href{https://www.ircad.fr/research/data-sets/}{2010}) & 1(FR) & L, SV, GB, AO, LES & \cmark & expert 
    \\
    
    VISCERAL (\href{https://visceral.eu/benchmarks/}{2015}) \cite{viserectal} & -- & L, GB, P, SP, K & -- & expert 
    \\
    
    BTCV (\href{https://www.synapse.org/Synapse:syn3193805/wiki/217789}{2015}) \cite{btcv2015} & 1(US) & L, SV, GB, ST, P, SP, K, AG, AO, ES & -- & expert 
    \\
    
    Colorectal-Liver-Metastases (\href{https://www.cancerimagingarchive.net/collection/colorectal-liver-metastases/}{2017}) \cite{simpson2024preoperative} & 1(US) & L, HV, SV, LES & \cmark & expert 
    \\
    
    DenseVNet (\href{https://zenodo.org/records/1169361}{2018}) \cite{densevnet} & 2(US) & UAO & -- & expert 
    \\
    
    LiTS (\href{https://academictorrents.com/details/27772adef6f563a1ecc0ae19a528b956e6c803ce}{2018}) \cite{Lits2017} & 7(DE, NL, CA, IL, FR) & L, LES & \cmark & expert 
    \\
    
    MSD-CT - Spleen task (\href{http://medicaldecathlon.com/}{2018}) \cite{msd2022} & 1(US) & SP & \cmark & AI+expert 
    \\
    
    Pancreatic Cancer Survival Prediction (\href{https://wiki.cancerimagingarchive.net/pages/viewpage.action?pageId=37224869}{2018}) & 1(US) & P, LES & \cmark & expert 
    \\
    
    MSD-CT - Colon task (\href{http://medicaldecathlon.com/}{2018}) \cite{msd2022} & 1(US) & CO & \cmark & expert 
    \\
    
    SegThor (\href{https://competitions.codalab.org/competitions/21145}{2019}) \cite{segthor} & 1(FR) & AO, ES & -- & expert 
    \\
    
    CHAOS (\href{https://chaos.grand-challenge.org/Data/}{2019}) \cite{CHAOS2021} & 1(TR) & L & -- & expert 
    \\
    
    CT-ORG (\href{https://www.cancerimagingarchive.net/collection/ct-org/}{2020}) \cite{ctorg2020} & 8(DE, NL, CA, FR, IL, US) & L, K & \cmark & AI+expert 
    \\
    
    MSD-CT - Pancreas task (\href{http://medicaldecathlon.com/}{2020}) \cite{msd2022} & 1(US) & P, LES & \cmark & expert 
    \\
    
    MSD-CT - Liver task (\href{http://medicaldecathlon.com/}{2020}) \cite{msd2022} & 7(DE, NL, CA, IL, FR) & L & \cmark & expert 
    \\
    
    MSD-CT - HepaticVessel task (\href{http://medicaldecathlon.com/}{2020}) \cite{msd2022} & 1(US) & L, HV & \cmark & AI+expert 
    \\
    
    Pancreas-CT (\href{https://www.cancerimagingarchive.net/collection/pancreas-ct/}{2020})\cite{roth2016pancreasct} & 1(US) & P & -- & expert 
    \\
    
    AbdomenCT-1K (\href{https://doi.org/10.5281/zenodo.5903099}{2021})\cite{abdomenct2021} & 12(DE, NL, FR, IL, US, CA, CN) & L, P, SP, K, CT & \cmark & AI+expert 
    \\
    
    WORC - GIST dataset (\href{https://zenodo.org/records/5221034}{2021}) \cite{Starmans2021WORC} & 1(NL) & LES & \cmark & expert 
    \\
    
    WORC - CRLM dataset (\href{https://zenodo.org/records/5221034}{2021})\cite{worcdatabase} & 1(NL) & LES & \cmark & expert 
    \\
    
    Pediatric (\href{https://www.cancerimagingarchive.net/collection/pediatric-ct-seg/}{2022})\cite{pediatric} & 1(US) & UAO, AG, IN, CO, RE & -- & expert 
    \\
    
    WORD (\href{https://drive.google.com/drive/folders/1i2xbXxdEYnjNZVUtGZxYdwaeKmNmywnY}{2022}) \cite{word2022} &  1(CN) & UAO, AG, IN, CO, RE & -- & expert 
    \\
    
    AMOS (\href{https://zenodo.org/records/7262581}{2022})\cite{amos} & 2(CN) & UAO, AG, AO & -- & AI+expert 
    \\
    
    KiPA22 (\href{https://zenodo.org/records/6361938}{2022}) & 1(CN) & K, LES & \cmark & expert 
    \\
    
    StageII-Colorectal-CT (\href{https://www.cancerimagingarchive.net/collection/stageii-colorectal-ct/}{2022}) \cite{stageii_crc} & 1(CN) & LN, LES & \cmark & expert 
    \\
    
    HCC-TACE-Seg (\href{https://www.cancerimagingarchive.net/collection/hcc-tace-seg/}{2022}) \cite{moawad2021hcc} & 1(US) & L, LES & \cmark & expert 
    \\
    
    AutoPET (FDG-PET/CT) (\href{https://www.cancerimagingarchive.net/collection/fdg-pet-ct-lesions/}{2022}) \cite{automatedpet} & 2(DE) & LES & \cmark & expert 
    \\
    
    DAP Atlas (\href{https://www.synapse.org/Synapse:syn52287632.1/datasets/}{2023}) \cite{dapatlas2023} & NA(DE) & SV, AG, IN, CO, RE, AO, MES & \cmark & AI+expert 
    \\
    
    Abdominal Trauma Det (\href{https://www.rsna.org/rsnai/ai-image-challenge/abdominal-trauma-detection-ai-challenge}{2024}) \cite{rudie2024rsna} & 23(More than 10 countries\footnotemark[7]) & UAO, SV, IN, CO, RE, AO, MES & \cmark & AI+expert 
    \\
    
    TotalSegmentator (\href{https://zenodo.org/records/10047292}{2023}) \cite{totalsegmentator} & 8(CH) & SV, AG, IN, CO, AO & \cmark & AI+expert 
    \\

    AbdomenAtlas 1.1 (\href{https://www.zongweiz.com/dataset}{2024}) \cite{abdomenatlas2024} & 112(More than 10 countries\footnotemark[8]) & UAO, HV, SV, AG, IN, CO, RE, AO, CT & \cmark & AI+expert 
    \\
    
    FLARE23 (\href{https://codalab.lisn.upsaclay.fr/competitions/12239}{2023}) \cite{flare23} & 44(CN, CA, BR, US, DE, FR, IL, PL, UK) & UAO, AG, AO & \cmark & AI+expert 
    \\
    
    
    KiTS (\href{https://kits-challenge.org/kits23/}{2019}) \cite{kits2019} & 1(US) & K, LES & \cmark & expert 
    \\
    
    CPTAC-PDA-Tumor-Annotations (\href{https://www.cancerimagingarchive.net/analysis-result/cptac-pda-tumor-annotations/}{2023})\cite{rozenfeld2023cptac} & NA & LN, LES & \cmark & AI+expert 
    \\
    
    CPTAC-CCRCC-Tumor-Annotations (\href{https://wiki.cancerimagingarchive.net/pages/viewpage.action?pageId=157288300}{2023}) \cite{clark2013tcia} & NA & LN, LES & \cmark & AI+expert 
    \\
    
    CARE (\href{https://paperswithcode.com/dataset/care}{2023}) \cite{zhang2023care} & 1(CN) & RE, LES & \cmark & expert 
    \\
    
    Low-dose (\href{https://www.cancerimagingarchive.net/collection/ldct-and-projection-data/}{2023})\cite{moen2021lowdose} & 2(US) & LES & \cmark & expert 
    \\
    
    SEG.A. (\href{https://multicenteraorta.grand-challenge.org/}{2023}) & NA & AO & -- & -- 
    \\
    
    CT Lymph Nodes (\href{https://www.cancerimagingarchive.net/collection/ct-lymph-nodes/}{2023})\cite{roth2014lymphnode} & 1(US) & LN & -- & expert 
    \\
    
    Adrenal-ACC-Ki67-Seg (\href{https://www.cancerimagingarchive.net/collection/adrenal-acc-ki67-seg/}{2023})\cite{ahmed2020radiomic} & 1(US) & LES & \cmark & AI+expert 
    \\
    
    AIMI Annotations Initiative (\href{https://zenodo.org/records/13244892}{2024}) \cite{vanoss2024aimi} & NA & L, K, LES & \cmark & AI + Randomly Revised 
    \\
    
    CURVAS (\href{https://zenodo.org/records/10979642}{2024}) \cite{riera2024curvas} & 1(DE) & L, P, K & \cmark & AI+expert 
    \\
    \hline
    \end{tabular}
\end{table*}

\footnotetext[1]{number of accessible and private cases}

\footnotetext[2]{US: United States, DE: Germany, NL: Netherlands, CA: Canada, IL: Israel, FR: France, TR: Turkey, CN: China, CH: Switzerland, MT: Malta, IE: Ireland, BR: Brazil, BA: Bosnia and Herzegovina, AUS: Australia, TH: Thailand, TW: Taiwan, CL: Chile, MA: Morocco, ES: Spain, PL: Poland, UK: United Kingdom}

\footnotetext[3]{ UAO: Upper Abdominal Organs(L, GB, SP, P, K, DU, ES, ST), L: Liver, HV: Hepatic Vessel, PV: Portal Vein, SV: Splenic Vein, PSV: Portal and Splenic Veins, GB: Gallbladder, ST: Stomach, P: Pancreas, SP: Spleen, K: Kidney, AG: Adrenal Gland, DU: Duodenum, IN: Intestine, CO: Colon, RE: Rectum, AO: Aorta, IVC: Inferior Vena Cava, CT: Celiac Trunk, MES: Mesentery, LN: Lymph Node, ES: Esophagus, LES: Lesion}

\footnotetext[4]{LiTS\cite{Lits2017}, KiTS\cite{kits2019}, Pancreas-CT\cite{roth2016pancreasct}, MSD-CT - pancreas and spleen tasks\cite{msd2022}}

\footnotetext[5]{
CHAOS\cite{CHAOS2021}, BTCV\cite{btcv2015}, CT-ORG\cite{ctorg2020}, Pancreas-CT \cite{roth2016pancreasct}, WORD\cite{word2022}, LiTS\cite{Lits2017}, AMOS\cite{amos}, KiTS\cite{kits2019}, AbdomenCT-1K\cite{abdomenct2021}, MSD-CT -  Pancreas, Spleen, Liver, Hepatic Vessel and Colon tasks\cite{msd2022}, Abdominal Trauma Det\cite{rudie2024rsna}, FLARE23\cite{flare23}, DAP Atlas\cite{dapatlas2023}, TotalSegmentator\cite{totalsegmentator}, AutoPET (FDG-PET/CT) \cite{automatedpet}
}

\footnotetext[6]{
LiTS\cite{Lits2017},KiTS\cite{kits2019}, MSD-CT - Pancreas, Spleen, Hepatic Vessel tasks\cite{msd2022},CPTAC-CCRCC-Tumor-Annotations\cite{clark2013tcia}, HCC-TACE-Seg\cite{moawad2021hcc}, DAP Atlas\cite{dapatlas2023}, StageII-Colorectal-CT\cite{stageii_crc}, AMOS\cite{amos}, WORD\cite{word2022}, CT Lymph Nodes\cite{roth2014lymphnode}
}

\footnotetext[7]{
countries: AUS, BA, CA, CL, DE, IE, MT, MA, ES, TH, TW, TR, US, BR }

\footnotetext[8]{
countries: MT, IE, BR, BA, AUS, TH, TW, CA, TR, CL, ES, MA, US, DE, NL, FR, IL, CN, CH}

\subsection{Dataset Composition and Redundancy}

\begin{figure}[hbtp!]
    \centering
    \includegraphics[width=1\columnwidth]
    {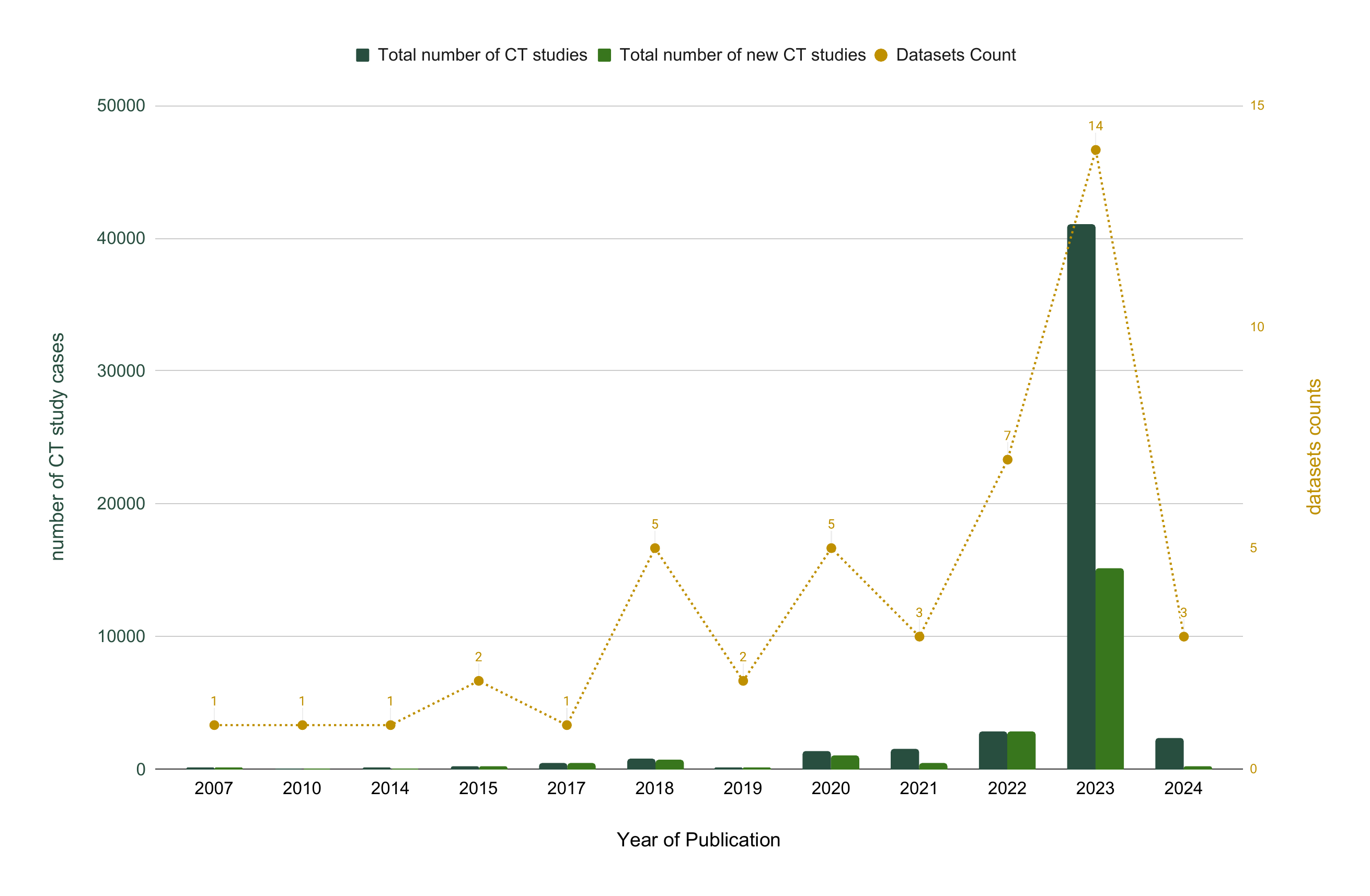} 
    \caption{Publication of datasets (line) trend and the number of new CT studies instances (bars) from 2007 to 2024.}
    \label{fig:newtrendchart}
\end{figure}

Figure \ref{fig:newtrendchart} highlights a notable trend in dataset composition—the frequent reuse of the same CT studies across multiple datasets. While this practice can improve resource efficiency, it also reduces data diversity and may hinder the generalizability of AI models to varied clinical scenarios. Among the 50,256 CT studies examined, only 20,559 were unique, revealing substantial redundancy. Such overlap raises the risk of data leakage during model training, which can lead to overfitting, where models learn to memorize specific cases rather than develop robust, generalizable representations.

\subsection{Geographic Distribution of Datasets}

Publicly available abdominal CT datasets have predominantly been acquired using scanners from major manufacturers such as GE, Siemens, Philips, and Toshiba, with 16- and 64-detector configurations being the most frequently reported systems [cite]. Despite contributions from 18 countries, the geographic distribution of datasets is heavily imbalanced, with a clear overrepresentation of high-income regions.

Approximately 75\% of the datasets originate from the United States, Canada, and European countries, reflecting a strong Western bias. The United States alone accounts for 21\% of all datasets, making it the most prolific single contributor, followed by China and France, each contributing 9\%. In contrast, datasets from non-Western regions—including Turkey, Taiwan, Chile, Morocco, Bosnia, and Brazil—collectively represent only 22\% of the total, indicating a substantial underrepresentation of low- and middle-income countries.

Notably, several major global regions—such as most of Africa, South Asia, and the Middle East—are entirely absent from the current dataset landscape. This geographic concentration limits the diversity of imaging sources and raises critical concerns about the generalizability of AI models trained on these datasets. Models developed from such regionally skewed data may underperform when applied in underrepresented healthcare settings, ultimately hindering equitable deployment and clinical utility across global populations.

\subsection{Organ and Pathology Distribution}

\begin{figure}[hbtp!]
    \centering
    \includegraphics[width=1\columnwidth]
    {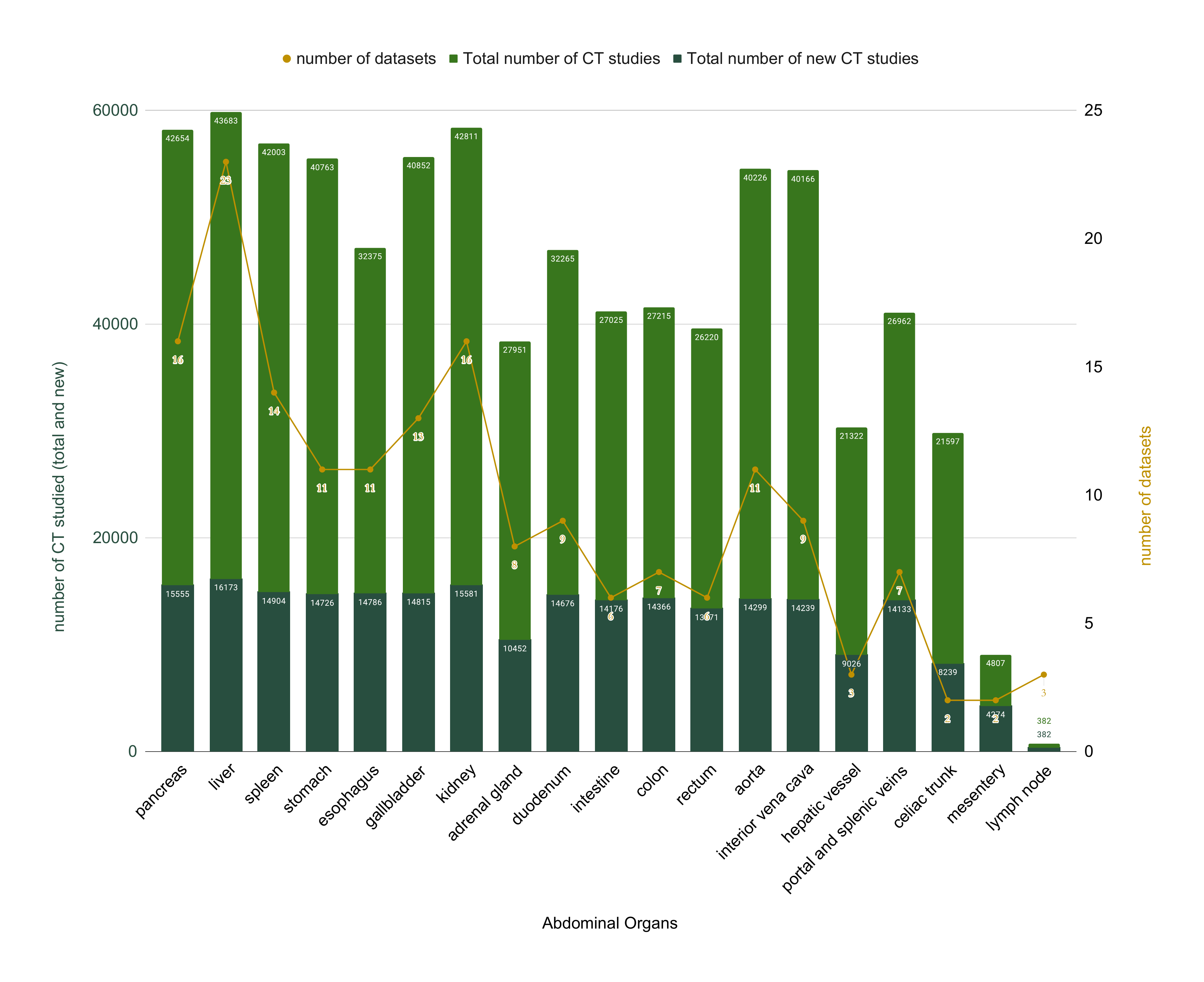}
    \caption{Abdominal organs concentration in datasets}
    \label{fig:abdorganbarchart}
\end{figure}

Figure \ref{fig:abdorganbarchart} and Table \ref{tab:pathology_organs} summarize the distribution of organ-specific abnormalities across publicly and privately available abdominal CT datasets. The liver and pancreas are the most frequently annotated organs, with rich datasets available for both healthy and diseased states. Liver pathologies span a broad clinical spectrum, including primary hepatic tumors, metastases, trauma-related injuries, and post-treatment imaging—highlighting the liver’s prominence in abdominal imaging research. Likewise, pancreatic datasets frequently include cases of cystic lesions and malignancies, reflecting the organ's diagnostic complexity and clinical importance.

Abnormalities in the kidneys and spleen are also well-documented, especially in the context of neoplasms, cysts, and trauma. In contrast, although imaging data for the gallbladder and adrenal glands are present in several datasets, the frequency of labeled abnormalities for these organs is markedly lower. This discrepancy may reflect either a lower incidence of clinically significant findings or a lack of detailed annotation in existing resources.

Beyond solid organs, several datasets include colorectal, rectal, and other gastrointestinal lesions, emphasizing the relevance of abdominal CT imaging in oncology applications. Despite this breadth, notable gaps remain. Many common, non-neoplastic conditions—such as inflammatory or vascular diseases—are underrepresented, which may inadvertently bias AI models toward tumor-centric tasks. As a result, these models risk underperforming in more general diagnostic scenarios, limiting their utility in routine clinical practice.

Addressing this imbalance will require broader annotation efforts and the inclusion of diverse pathologies to ensure AI tools are developed with a more comprehensive diagnostic foundation.

\begin{table*}[t]
    \centering
    \caption{Public and Private CT Counts with Anomaly Types per Organ} 
    \label{tab:pathology_organs}
    \footnotesize
    \setlength{\tabcolsep}{4pt}
    \renewcommand{\arraystretch}{1.1}
    \begin{tabular}{|l|r|r|P{0.58\textwidth}|} 
        \hline
        \textbf{Organ} & \textbf{Public} & \textbf{Private} & \textbf{Anomaly Type - Public+Private count} \\
        \hline

        Pancreas & 16206 & 12640 & Pancreas cyst or cancer (579+245) \\
        
        Liver & 16663 & 13541 & Mix of pre- and post-therapy images of primary and metastatic tumors (111+70), \newline Mix of benign and malignant lesions (most of 119+21), \newline Liver tumor (252), FNH (2), HCC pre-TACE (105) and post-TACE (105), \newline Colorectal liver metastasis pre- (197) and post-procedure (197), \newline Traumatic liver injury (340+151) \\
        
        Spleen & 15825 & 13551 & Spleen injury (372+145) \\
        
        Stomach & 18028 & 13551 & Bowel (71+62) and mesenteric injury and active extravasation (215+121) \\
        
        Esophagus & 8769 & 10318 & Bowel (71+62) and mesenteric injury and active extravasation (215+121) \\
        
        Gallbladder & 14722 & 12828 & -- \\
        
        Kidney & 16440 & 13782 & Kidney tumor (1088+30), \newline Cyst (268+70), \newline Injury (217+153) \\
        
        Adrenal gland & 9293 & 9575 & -- \\
        
        Duodenum & 9250 & 10278 & Bowel (71+62) and mesenteric injury and active extravasation (215+121) \\
        
        Intestine & 11400 & 10178 & Bowel (71+62) and mesenteric injury and active extravasation (215+121) \\
        
        Colon & 11401 & 10178 & Primary colon cancer (126+64), \newline Bowel (71+62) and mesenteric injury and active extravasation (215+121) \\
        
        Rectum & 11797 & 10178 & Rectal cancer (436), \newline Bowel (71+62) and mesenteric injury and active extravasation (215+121) \\
        
        Aorta & 14188 & 12828 & Different aortic pathologies in some of 56 cases \\
        
        IVC & 14092 & 12828 & -- \\
        
        Hepatic Vessels & 9676 & 11223 & Mix of primary and metastatic liver tumors (303+140), \newline Colorectal liver metastasis pre- (197) and post-procedure (197) \\
        
        Portal and Splenic Veins & 11441 & 9475 & Colorectal liver metastasis pre- (197) and post-procedure (197) \\
        
        Celiac Trunk & 9374 & 11223 & -- \\
        
        Mesenteric Vessels & 4084 & 732 & Mesenteric injury and active extravasation (215+121) \\
        
        Lymph Nodes & 338 & 0 & Peritumoral lymph nodes in colorectal cancer (230), \newline Lymphadenopathy in pancreas (10) and colorectal cancer (12), \newline Non-cancerous lymphadenopathy (86) \\
        
        Abdominal Lesion & 4212 & 442 & 
        General: malignant lymphoma, melanoma, and non-small cell lung cancer (501+150). \newline
        Gastrointestinal (Public): GIST (126) and other pathologies mimicking GIST, including: \newline
        - Schwannoma (22), \newline
        - Leiomyosarcoma (25), \newline
        - Esophageal or GEJ carcinoma (25), \newline
        - Lymphoma (25). \newline
        Colorectal: Colorectal Cancer Stage II (230+0), Rectal cancer (399+0). \newline
        Liver (Public): Metastasis (314), post-procedure Colorectal Metastasis (197), \newline
        Tumor (120), post-TACE HCC (105), Cyst (26), Hemangioma (4), \newline
        Focal fat/perfusion (4), FNH (2), Ablation defect (1). \newline
        Liver (Public+Private): pre- and post-therapy images of primary and metastatic tumors (194). \newline
        Kidney: Kidney tumors (1143+30) and cysts (248+NA). \newline
        Pancreas: Pancreas Cancer (249+53), Pancreatic cyst or tumor (281+139). \newline
        Adrenal: Adrenocortical carcinoma (53+0). \\        
        \hline
    \end{tabular}
\end{table*}

\subsection{Annotation Practices and Dataset Bias}
Annotation methodologies vary across datasets, impacting the reliability of AI model training:
\begin{itemize}
    \item 60\% of datasets rely on manual annotation by radiologists and trained experts.
    \item 35\% use AI-assisted labeling, where AI-generated annotations are later refined by human experts.
    \item 5\% are fully AI-annotated, introducing potential concerns regarding labeling accuracy.
\end{itemize}

While AI-assisted annotation offers efficiency gains, studies suggest that fully AI-generated labels may introduce systematic errors, particularly in complex segmentation tasks \cite{Kocak2024}. Ensuring annotation consistency across datasets is crucial for reliable AI training.

\subsection{The bias evaluation revealed substantial disparities in dataset fairness} 

\begin{figure}[hbtp!]
    \centering
    \includegraphics[width=1.0\columnwidth]
    {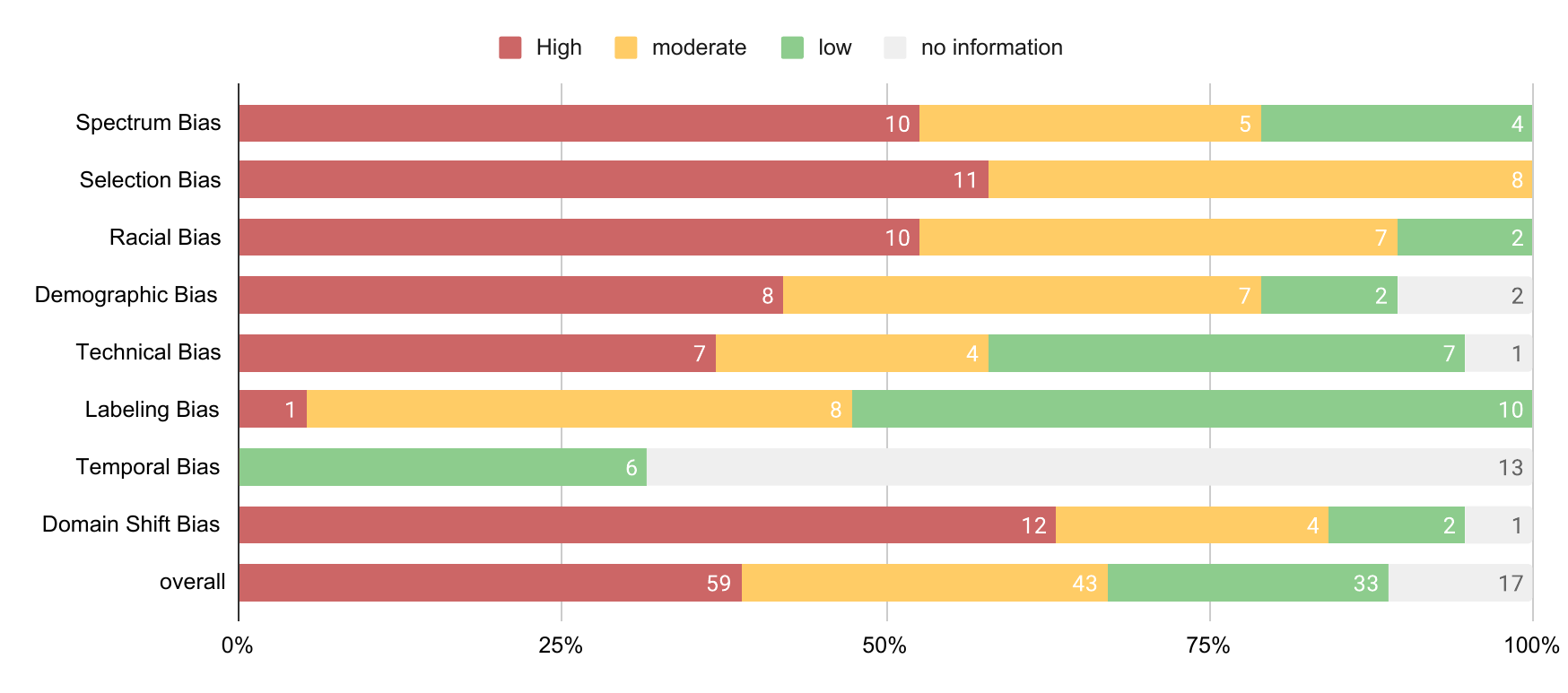} 
    \caption{Datasets bias in scale of high, moderate, and low}
    \label{fig:generic_bar}
\end{figure}

Table~\ref{tab:rob-assessment} summarizes the bias evaluation, which excluded datasets containing fewer than 100 cases, resulting in a final set of 19 datasets. Figure~\ref{fig:generic_bar} shows that the most common bias types were domain shift bias (63\%) and selection bias (57\%), indicating that many datasets may not generalize effectively beyond their original clinical environments. Spectrum bias (52\%) and racial bias (52\%) were also prevalent, suggesting over-representation of specific pathologies and patient demographics, which may adversely affect model fairness. Labeling bias was least frequent (5.3\%), implying that annotation inconsistencies are a comparatively minor concern relative to dataset composition. In total, 47\% of datasets (n = 9) exhibited three or more high-risk bias indicators, whereas only 10\% (n = 2) had three or more low-risk ratings. These findings demonstrate significant disparities in fairness and representational validity, underscoring the need for more diverse and balanced datasets to improve AI model generalizability and fairness.
 
\begin{table}[t]
\caption{Risk of bias assessment for datasets with \(\geq\)100 cases (\(n=19\))}

\footnotesize
\setlength{\tabcolsep}{2pt}       
\renewcommand{\arraystretch}{1.05}

\begin{adjustbox}{max width=\columnwidth}
\begin{tabularx}{\columnwidth}{|>{\raggedright\arraybackslash}X|
                              *{8}{Y|}}
\hline
\thead{Dataset/\\RoB} & {Spec.} & {Sel.}
& {Race.} & {Demo.} & {Tech.}
& {Lbl} & {Temp.} & {D-Shift} \\
\hline
        MSD-CT - Pancreas task(\href{http://medicaldecathlon.com/}{2020}) \cite{msd2022} & \cellcolor{high}H & \cellcolor{high}H & \cellcolor{moderate}M & \cellcolor{moderate}M & \cellcolor{moderate}M & \cellcolor{moderate}M & \cellcolor{low}L & \cellcolor{high}H \\
        
        LiTS(\href{https://academictorrents.com/details/27772adef6f563a1ecc0ae19a528b956e6c803ce}{2018}) \cite{Lits2017} & \cellcolor{moderate}M & \cellcolor{moderate}M & \cellcolor{low}L & \cellcolor{moderate}M & \cellcolor{low}L & \cellcolor{low}L & \cellcolor{low}L & \cellcolor{low}L \\
        
        AutoPET(\href{https://www.cancerimagingarchive.net/collection/fdg-pet-ct-lesions/}{2022}) \cite{automatedpet} & \cellcolor{low}L & \cellcolor{moderate}M & \cellcolor{moderate}M & \cellcolor{moderate}M & \cellcolor{high}H & \cellcolor{moderate}M & \cellcolor{noinfo}NA & \cellcolor{moderate}M \\
        
        AMOS(\href{https://zenodo.org/records/7262581}{2022})\cite{amos} & \cellcolor{moderate}M & \cellcolor{moderate}M & \cellcolor{high}H & \cellcolor{moderate}M & \cellcolor{low}L & \cellcolor{low}L & \cellcolor{noinfo}NA & \cellcolor{high}H \\
        
        WORD(\href{https://drive.google.com/drive/folders/1i2xbXxdEYnjNZVUtGZxYdwaeKmNmywnY}{2022}) \cite{word2022} & \cellcolor{moderate}M & \cellcolor{high}H & \cellcolor{high}H & \cellcolor{high}H & \cellcolor{high}H & \cellcolor{low}L & \cellcolor{noinfo}NA & \cellcolor{high}H \\
        
        TotalSegmentator(\href{https://zenodo.org/records/10047292}{2023}) \cite{totalsegmentator} & \cellcolor{low}L & \cellcolor{moderate}M & \cellcolor{high}H & \cellcolor{moderate}M & \cellcolor{low}L & \cellcolor{moderate}M & \cellcolor{low}L & \cellcolor{high}H \\
        
        AbdomenAtlas(\href{https://www.zongweiz.com/dataset}{2024}) \cite{abdomenatlas2024} & \cellcolor{moderate}M & \cellcolor{moderate}M & \cellcolor{moderate}M & \cellcolor{low}L & \cellcolor{low}L & \cellcolor{low}L & \cellcolor{noinfo}NA & \cellcolor{moderate}M \\
        
        Abdominal Trauma Det(\href{https://www.rsna.org/rsnai/ai-image-challenge/abdominal-trauma-detection-ai-challenge}{2024}) \cite{rudie2024rsna} & \cellcolor{low}L & \cellcolor{moderate}M & \cellcolor{low}L & \cellcolor{low}L & \cellcolor{low}L & \cellcolor{low}L & \cellcolor{noinfo}NA & \cellcolor{moderate}M \\
        
         KiTS (\href{https://kits-challenge.org/kits23/}{2019}) \cite{kits2019} & \cellcolor{high}H & \cellcolor{high}H & \cellcolor{high}H & \cellcolor{high}H & \cellcolor{moderate}M & \cellcolor{moderate}M & \cellcolor{low}L & \cellcolor{high}H \\
         
        MSD-CT - HepaticVessel task(\href{http://medicaldecathlon.com/}{2020}) \cite{msd2022} & \cellcolor{high}H & \cellcolor{high}H & \cellcolor{high}H & \cellcolor{high}H & \cellcolor{moderate}M & \cellcolor{moderate}M & \cellcolor{noinfo}NA & \cellcolor{high}H \\
        
        Pediatric(\href{https://www.cancerimagingarchive.net/collection/pediatric-ct-seg/}{2022})\cite{pediatric} & \cellcolor{low}L & \cellcolor{moderate}M & \cellcolor{moderate}M & \cellcolor{high}H & \cellcolor{low}L & \cellcolor{low}L & \cellcolor{noinfo}NA & \cellcolor{moderate}M \\
        
        KiPA(\href{https://zenodo.org/records/6361938}{2022}) & \cellcolor{high}H & \cellcolor{high}H & \cellcolor{high}H & \cellcolor{high}H & \cellcolor{high}H & \cellcolor{low}L & \cellcolor{low}L & \cellcolor{high}H \\
        
        HCC-TACE-Seg(\href{https://www.cancerimagingarchive.net/collection/hcc-tace-seg/}{2022}) \cite{moawad2021hcc} & \cellcolor{high}H & \cellcolor{high}H & \cellcolor{moderate}M & \cellcolor{moderate}M & \cellcolor{high}H & \cellcolor{low}L & \cellcolor{noinfo}NA & \cellcolor{high}H \\
        
        Colorectal-Liver-Metastases(\href{https://www.cancerimagingarchive.net/collection/colorectal-liver-metastases/}{2017}) \cite{simpson2024preoperative} & \cellcolor{high}H & \cellcolor{high}H & \cellcolor{moderate}M & \cellcolor{high}H & \cellcolor{high}H & \cellcolor{moderate}M & \cellcolor{noinfo}NA & \cellcolor{high}H \\
        
        PanCan Survival Prediction(\href{https://wiki.cancerimagingarchive.net/pages/viewpage.action?pageId=37224869}{2018}) & \cellcolor{high}H & \cellcolor{high}H & \cellcolor{moderate}M & \cellcolor{noinfo}NA & \cellcolor{noinfo}NA & \cellcolor{moderate}M & \cellcolor{noinfo}NA & \cellcolor{noinfo}NA \\
        CARE(\href{https://paperswithcode.com/dataset/care}{2023}) \cite{zhang2023care} & \cellcolor{high}H & \cellcolor{high}H & \cellcolor{high}H & \cellcolor{noinfo}NA & \cellcolor{high}H & \cellcolor{low}L & \cellcolor{noinfo}NA & \cellcolor{high}H \\
        
        WORC(\href{https://zenodo.org/records/5221034}{2021}) \cite{Starmans2021WORC} & \cellcolor{moderate}M & \cellcolor{moderate}M & \cellcolor{high}H & \cellcolor{moderate}M & \cellcolor{low}L & \cellcolor{moderate}M & \cellcolor{low}L & \cellcolor{low}L \\
    \hline

\end{tabularx}
\end{adjustbox}

\vspace{2pt}
\emph{\footnotesize
Abbrev.: RoB=Risk of Bias; Spec.=Spectrum; Sel.=Selection; Race.=Racial;
Demo.=Demographic; Tech.=Technical; Lbl.=Labeling; Temp.=Temporal; D-Shift=Domain Shift.
H/M/L = high/moderate/low risk; 
}

\label{tab:rob-assessment}
\end{table}

Table~\ref{tab:top_datasets} presents the top five datasets identified as most reliable for AI model training, selected based on dataset size, annotation method, organ coverage, and bias indicators. These datasets contain a large number of annotated volumes, provide comprehensive organ coverage, and demonstrate low bias risk across key evaluation metrics. Most annotations were performed by expert radiologists, ensuring high labeling quality. Overall, these datasets offer well-balanced and diverse samples, increasing the likelihood that AI models trained on them will generalize effectively across varied clinical scenarios.

\begin{table}[t]
    \centering
    \caption{Selected Abdominal Imaging Datasets with Maximum Case Coverage and Minimal Bias}
    \footnotesize
    \setlength{\tabcolsep}{4pt}
    \renewcommand{\arraystretch}{1.1}
    \resizebox{\columnwidth}{!}{ 
    \begin{tabular}{|P{0.22\columnwidth}|P{0.08\columnwidth}|P{0.14\columnwidth}|P{0.13\columnwidth}|P{0.43\columnwidth}|}


        \hline
        \textbf{Dataset} & \textbf{Cases} & \textbf{Organs} & \textbf{Bias Risk} & \textbf{Subjects status} \\
        \hline
        
        LiTS \cite{Lits2017} & 201 & Liver, Lesions & Low & Liver cancer \newline(pre/post-therapy) \\
        
        AutoPET \cite{automatedpet} & 1164 & Lesions & Moderate & Oncological cases \newline(NSCLC, lymphoma, melanoma) \\
        
        AbdAtlas \cite{abdomenatlas2024} & 9262 & 16 Organs & Low & Normal/cancerous organs \newline(colorectal, pancreatic) \\
        
        AbdTrauma \cite{rudie2024rsna} & 3551 & 14 Organs & Moderate & Traumatic injuries \\
        
        Pediatric \cite{pediatric} & 359 & 12 Organs & Low & Pediatric cases\\
        
        WORC \cite{Starmans2021WORC} & 323 & Tumors, Lesions & Moderate & GIST\newline intra-abdominal tumors \\
        \hline
    \end{tabular}
    }
    \label{tab:top_datasets}
\end{table}

\section{Discussion}

Table~\ref{tab:datasets_1} shows that dataset redundancy is a prominent issue, with 59\% of CT studies reused across multiple datasets, thereby reducing diversity and increasing the risk of data leakage. Figure~\ref{fig:generic_bar} further illustrates that current resources exhibit limited representation of the real-world population, with insufficient variation in disease spectrum, demographic diversity, and standardized labeling practices. Table~\ref{tab:rob-assessment} indicates that 47\% of datasets demonstrate high-risk bias in three or more categories. These findings highlight both the potential and the limitations of the current abdominal CT dataset landscape, where rapid advances in imaging and AI are counterbalanced by persistent technical and ethical challenges.

Figure~\ref{fig:generic_bar} also shows that domain shift bias affects 63\% of datasets, underscoring the complexity of sharing and integrating multi-center data. While multi-institutional collaboration can yield larger and more representative datasets, differences in scanner models, contrast protocols, and labeling standards introduce systematic variability. Regional variations in disease prevalence, genetic factors, and clinical workflows further limit dataset generalizability. Even advanced methods such as domain adaptation or batch-effect correction may be insufficient to mitigate these biases, making heterogeneous data sources a persistent barrier to developing robust models for diverse healthcare settings.

Figure~\ref{fig:abdorganbarchart} shows that pancreas, followed by liver, spleen, stomach, and kidney datasets, are the most prevalent in abdominal imaging research. In contrast, organs such as the esophagus, adrenal glands, and lymph nodes are markedly underrepresented, indicating a substantial imbalance in organ coverage across available datasets. Table~\ref{tab:pathology_organs} further reveals that most datasets focus on tumor detection, potentially limiting model applicability to a broader range of pathologies. Although these imbalances constrain generalizability, current datasets remain valuable for algorithm development and proof-of-concept studies. They enable researchers to rapidly prototype and refine methods before clinical deployment. Emerging technologies, including large language models (LLMs) and foundation AI systems, may further accelerate automated labeling and segmentation, enhancing pre-validation workflows.

Table~\ref{tab:top_datasets} identifies five datasets with balanced organ coverage and low bias risk, illustrating the importance of diversity and expert annotation in producing generalizable models. Federated collaboration, supported by standardized acquisition and labeling protocols, offers a path toward constructing datasets that reflect real-world disease distributions. Ethical considerations—including patient consent and privacy—must be integrated at every stage. Both commercial and non-profit contributors should ensure compliance with relevant regulations while enabling scientific advancement.

Figure~\ref{fig:generic_bar} also highlights the dual role of emerging AI technologies: while they can streamline annotation, they may perpetuate existing biases if not subject to iterative quality control. Anchoring automated processes in continuous evaluation will be essential to creating datasets that serve as a gold standard for research and clinical use. Ultimately, robust curation, harmonization, and governance will be central to ensuring that abdominal CT datasets drive both innovation and equitable healthcare outcomes.

\section{Conclusion}
In conclusion, the current trajectory of abdominal CT datasets reveals a landscape rich in opportunity, yet challenged by variability, ethical imperatives, and the dynamic evolution of AI-driven methodologies. High-quality, multi-center data encompassing an expansive range of diseases, patient demographics, and imaging conditions are indispensable for building broadly applicable foundation models. However, local adaptation and fine-tuning of such models will likely remain a cornerstone for optimizing clinical relevance, given the substantial inter-center differences in patient characteristics and imaging protocols. As labeling tasks and segmentation processes become increasingly automated through the integration of LLMs and foundation models, systematic checks and rigorous evaluation of biases must be embedded within the research pipeline. Ethical considerations, including proper licensing and governance, will continue to shape how these data are collected, shared, and utilized. By thoughtfully balancing these elements, the field can evolve beyond the current limitations, harnessing the power of innovation to create robust, equitable, and clinically impactful abdominal CT datasets.

\section*{Data and Materials Availability}
The full extraction sheet (dataset inventory, fields, and bias ratings) is publicly available at Google Sheet link: 
\url{https://docs.google.com/spreadsheets/d/1l_2GHLyl3zAB_Eb3veYtj6PLBA1hXMaNgG8Z2q9eu8Y/edit?usp=sharing}.

\end{document}